\documentclass[sigconf,natbib=true,anonymous=false]{acmart}

\usepackage{amsfonts}
\usepackage{amssymb}
\usepackage{amsthm}
\usepackage{multirow}
\usepackage{caption}
\usepackage{subcaption}
\newtheorem{theorem}{Theorem}

\AtBeginDocument{%
  \providecommand\BibTeX{{%
    \normalfont B\kern-0.5em{\scshape i\kern-0.25em b}\kern-0.8em\TeX}}}

\copyrightyear{2023}
\acmYear{2023}
\setcopyright{acmlicensed}
\acmConference[SIGIR '23] {Proceedings of the 46th International ACM SIGIR Conference on Research and Development in Information Retrieval}{July 23--27, 2023}{Taipei, Taiwan.}
\acmBooktitle{Proceedings of the 46th International ACM SIGIR Conference on Research and Development in Information Retrieval (SIGIR '23), July 23--27, 2023, Taipei, Taiwan}
\acmPrice{15.00}
\acmISBN{978-1-4503-9408-6/23/07}
\acmDOI{10.1145/XXXXXX.XXXXXX}

\begin{document}

\title{Intent-aware Ranking Ensemble for Personalized Recommendation}

\author{Jiayu Li}
\email{jy-li20@mails.tsinghua.edu.cn}
\affiliation{
  \institution{ DCST, Tsinghua University}
  \state{Beijing}
  \country{China}
}

\author{Peijie Sun}
\email{sun.hfut@gmail.com}
\affiliation{
  \institution{ DCST, Tsinghua University}
  \state{Beijing}
  \country{China}
}

\author{Zhefan Wang}
\email{wzf19@mails.tsinghua.edu.cn}
\affiliation{
  \institution{ DCST, Tsinghua University}
  \state{Beijing}
  \country{China}
}

\author{Weizhi Ma}
\email{mawz@tsinghua.edu.cn}
\affiliation{
  \institution{ AIR, Tsinghua University\\Beijing Academy of Artificial Intelligence}
  \state{Beijing}
  \country{China}
}

\author{Yangkun Li}
\email{liyangkun17@gmail.com}
\affiliation{
  \institution{ DCST, Tsinghua University}
  \state{Beijing}
  \country{China}
}

\author{Min Zhang}
\authornote{Min Zhang is the corresponding author.}
\email{z-m@tsinghua.edu.cn}
\affiliation{
  \institution{ DCST, Tsinghua University\\BNRist}
  \state{Beijing}
  \country{China}
}

\author{Zhoutian Feng}
\email{fengzhoutian@hotmail.com}
\affiliation{
  \institution{ Meituan Inc. }
  \state{Beijing}
  \country{China}
}

\author{Daiyue Xue}
\email{xuedaiyue@meituan.com}
\affiliation{
  \institution{ Meituan Inc. }
  \state{Beijing}
  \country{China}
}
\begin{abstract}
Ranking ensemble is a critical component in real recommender systems. 
When a user visits a platform, the system will prepare several item lists, each of which is generally from a single behavior objective recommendation model.
As multiple behavior intents, e.g., both clicking and buying some specific item category, are commonly concurrent in a user visit, it is necessary to integrate multiple single-objective ranking lists into one.
However, previous work on rank aggregation mainly focused on fusing homogeneous item lists with the same objective while ignoring ensemble of heterogeneous lists ranked with different objectives with various user intents.

In this paper, we treat a user's possible behaviors and the potential interacting item categories as the user's intent. And we aim to study how to fuse candidate item lists generated from different objectives aware of user intents. 
To address such a task, we propose an Intent-aware ranking Ensemble Learning~(IntEL) model to fuse multiple single-objective item lists with various user intents, in which item-level personalized weights are learned.
Furthermore, we theoretically prove the effectiveness of IntEL with point-wise, pair-wise, and list-wise loss functions via error-ambiguity decomposition. 
Experiments on two large-scale real-world datasets also show significant improvements of IntEL on multiple behavior objectives simultaneously compared to previous ranking ensemble models.

\end{abstract}


\maketitle

\section{Introduction}
\label{sec:intro}
\begin{figure}
    \centering
    \includegraphics[width=0.95\linewidth]{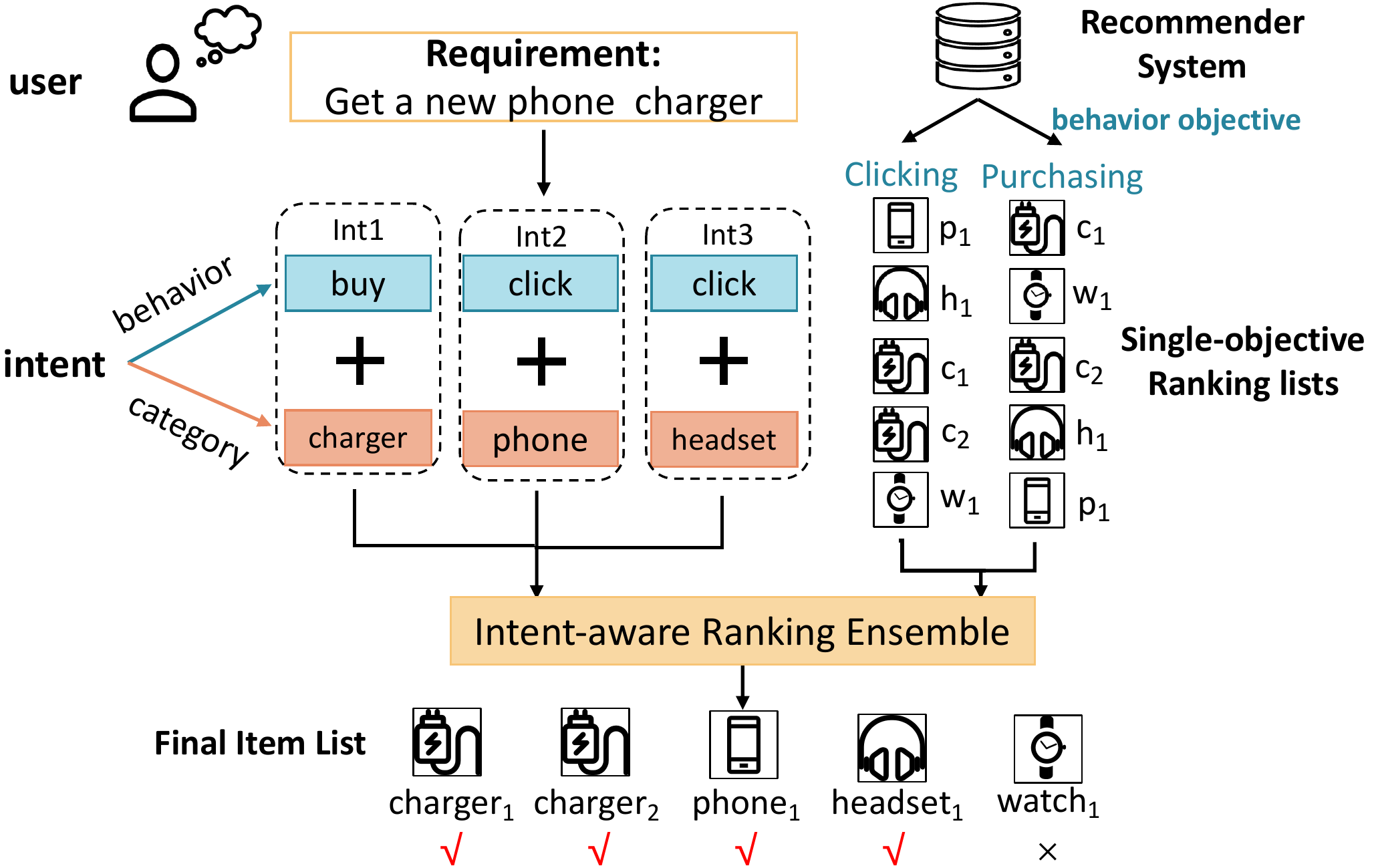}
    \caption{An example of fusing two single-objective item lists into a final list aware of multiple concurrent user intents.} 
    \label{fig:illustration}
\end{figure}

Users typically have various intents when using recommender systems.
For instance, when shopping online, users may intend to buy snacks or browse clothes. 
Generally, we call the users' behaviors the behavior intents and their interacted item categories the item category intents. 
Multiple behavior intents may be concurrent in a visit, and users need distinct items with different item category intents.
Therefore, user intents are essential to recommender systems for recommendation list generation. In this paper, we follow the definition of user intents by~\citet{chen2019air} as a combination of user behavior and item category, such as booking an item with a hotel category or clicking an item in a phone category.

From the systems' viewpoint, since users usually have diverse intents, multiple item lists will be generated when a user visits the platform. These lists generally come from recommendation models optimized with different behavior objectives, such as clicking, consuming, or browsing duration.
Existing research has made promising achievements with a single objective, such as predicting Click Through Rate~(CTR)~\cite{zhu2021open,guo2017deepfm,bian2022can} and Conversion Rate~(CVR)~\cite{pan2022metacvr,dai2022generalized}.
However, as multiple intents of a user may appear in a visit, it is crucial to aggregate multiple heterogeneous single-objective ranking lists aware of the user's current intents.

An example of the intent-aware ranking ensemble on an online shopping platform is shown in Figure~\ref{fig:illustration}.
A user needs to buy a phone charger, and she also wants to browse new products about phones and headsets.
The system has two single-objective ranking lists ready when she visits the platform. These lists are produced by two recommendation models optimized with users' consumption and clicking histories, respectively.
To satisfy the user's diverse intents at once, an intent-aware ranking ensemble model is adopted to aggregate two ranking lists for a final display, where items are reordered according to both basic ranking lists and the user's intents.
Thus, charger$_1$, charger$_2$, phone$_1$, and headset$_1$ are placed at the front of the final list, satisfying users' preference better than both single-objective ranking lists.
Therefore, intent-aware ranking ensemble is important for promoting recommendation performance.

However, there have been few attempts to combine heterogeneous single-objective ranking lists~(Hereinafter referred to as basic lists) considering user intents.
In industry, a common strategy is simply summing basic lists with pre-defined list-level weights, which ignores users' personalized preference.
While in academia, existing studies are not adequate to handle ranking ensemble for personalized recommendation.
Widely-explored unsupervised rank aggregation methods~\cite{bhowmik2017letor,kolde2012robust,liang2018manifold} are mostly studied in information retrieval tasks rather than recommendation scenario.
Recently, supervised methods~\cite{balchanowski2022collaborative,balchanowski2022aggregation,oliveira2016evolutionary} have been proposed to combine different item lists in recommendation.
Nevertheless, these studies focused on combining homogeneous item lists optimized for the same behavior, not the heterogeneous rank lists for different objectives. 
Users' intents are also overlooked in the ranking ensemble stage.

To aggregate basic lists aware of user intents, we aim to learn different weights for different basic lists and item categories to sum basic lists' scores.
However, it is challenging since numerous weights should be assigned for all items in all basic lists, which may be hard to learn.
Therefore, we first prove its effectiveness theoretically.
Unlike previous studies, we aim to assign ensemble weights at item level rather than list level.
We prove the effectiveness of this form of ranking ensemble and verify that the loss of the ensemble list can be smaller than the loss of any basic models with point-wise, pair-wise, and list-wise loss functions.
An ambiguity term is derived from the proof and used for optimization loss. 

With theoretical guarantees, another challenge in practice is to infer users' intents and integrate the intents into ranking ensemble of heterogeneous basic lists.
To address this challenge, we propose an Intent-aware ranking Ensemble Learning~(IntEL) method for personalized ensemble of multiple single-objective ranking lists adaptively. 
A sequential model is adopted to predict users' intents.
And a ranking ensemble module is designed to integrate basic list scores, item categories, and user intents.
Thus, the learnable ranking ensemble model can adaptively adjust the integration of multiple heterogeneous lists with user intents.

We conducted experiments on a public-available online shopping recommendation dataset and a local life service dataset. Our method, IntEL, is compared with various ensemble learning baselines and shows significant improvements.
The main contributions of this work are as follows:

\begin{itemize}
    \item To our knowledge, it is the first work that aims to generalize ranking ensemble learning with item-level weights on multiple heterogeneous item lists. We theoretically prove the effectiveness of ranking ensemble in this new setting.
    \item A novel intent-aware ranking ensemble learning model, IntEL, is proposed to fuse multiple single-objective recommendation lists aware of user intents adaptively. In the model, ambiguity loss, ranking loss, and intent loss have been proposed and integrated.
    \item Experiments on two large-scale real-world recommendation datasets indicate IntEL is significantly superior to previous ranking ensemble models on multiple objectives.
\end{itemize}

\section{Related Work}

\subsection{Ranking Ensemble}

Ranking ensemble, i.e., fusing multiple ranking lists for a consensus list, has been long discussed in IR scenario~\cite{dwork2001rank,farah2007outranking} and proved to be NP-hard even with small collections of basic lists~\cite{dwork2001rank2}.

In general, rank aggregation includes unsupervised and supervised methods.
Unsupervised methods only rely on the rankings.
For instance, Borda Count~\cite{borda1784memoire} computed the sum of all ranks. MRA~\cite{fagin2003efficient} adopted the median of rankings. 
Comparisons among basic ranks were also used, such as pair-wise similarity in Outrank~\cite{farah2007outranking}and distance from null ranks in RRA~\cite{kolde2012robust}.
Recently, researchers have paid attention to supervised rank aggregation methods.
For example, the Evolutionary Rank Aggregation~(ERA)~\cite{oliveira2016evolutionary} was optimized with genetic programming.
Differential Evolution algorithm~\cite{balchanowski2022aggregation,balchanowski2022collaborative} and reinforcement learning~\cite{zhang2022multi} were also adopted for rank aggregation optimization.
However, these rank aggregation methods only utilized the rank or scores of items in basic lists without considering item contents and users in recommendation.

Another view on the fusion problem comes from ensemble learning. It is a traditional topic in machine learning~\cite{sagi2018ensemble}, which has been successfully applied to various tasks~\cite{wang2014sentiment,verma2017comparative,mendes2012ensemble}.
A basic theory in ensemble learning is error-ambiguity~(EA) decomposition analysis~\cite{krogh1994neural}, which proves better performance can be achieved with aggregated results with good and diverse basic models.
It was proved in classification and regression with diverse loss functions~\cite{brown2005diversity,yin2014convex}.
\citet{liu2022generalized} generalized EA decomposition to model-level weights in ranking ensemble with list-wise loss, where different items in a list shared the same weights.

The differences between the previous studies and our method are mainly twofold:
First, rather than calculate a general weight for each basic model, we extend to assign item-level weights considering item category and user behavior intents.
We theoretically prove the effectiveness of this extension.
Second, 
we aim to combine heterogeneous lists generated for different behavior objectives and simultaneously improve performance on multiple objectives.

\subsection{Multi-Intent Recommendation}
Since we aggregate ranking lists aware of users' multiple intents, we briefly introduce recent methods on multi-intents and multi-interests in recommender systems.
Existing studies focused on capturing dynamic intents in the sequential recommendation~\cite{chen2019air,wang2020toward,liu2021intent,liu2020basket,wang2020intention}.
For instance, AIR~\cite{chen2019air} predicted intents and their migration in users' historical interactions.
\citet{wang2020toward} modeled users' dynamic implicit intention at the item level to capture item relationships.
MIND~\cite{sabour2017dynamic} and ComiRec~\cite{cen2020controllable} adopted dynamic routing from historical interactions to capture users' multi-intents and diversity.
TimiRec~\cite{wang2022target} distilled target user interest from predicted distribution on multi-interest of the users.
With the development of contrastive learning, implicit intent representations were also applied as constraints on contrastive loss~\cite{chen2022intent,di2022multi}.

Previous studies usually mixed ``intent'' and ``interest'' and paid attention to intent on item contents in single-behavior scenarios.
However, we follow~\cite{chen2019air} to consider both behavior intents and item category intents.
Moreover, instead of learning user preference for each intent, we utilize intents as guidance for fusing user preference with different behavior objectives.

\subsection{Multi-Objective Recommendation}

Another brunch of related but different work is the multi-objective recommendation.
It mainly contains two groups of studies.
One group provides multiple recommendation lists for different objectives with shared information among objectives, such as MMOE~\cite{ma2018modeling} and PLE~\cite{tang2020progressive}, where different lists are evaluated on corresponding objectives separately. 
The other group tried to promote the model performance on a target behavior objective with the help of other objectives, such as MB-STR~\cite{yuan2022multi} predicting users' click preferences.
However, instead of generating multiple lists or specifying a target behavior, we fuse a uniform list on which multiple objectives are evaluated simultaneously. 

Some studies that tried to jointly optimize ranking accuracy and other goals are also called multi-objective recommendation, such as fairness~\cite{wang2022survey}, diversity~\cite{castells2022novelty}, etc.
They sought to promote other metrics while maintaining utility on some behavior.
But we aim to concurrently promote performance on multiple objectives by aggregating various recommendation lists.

\section{Preliminaries}

\subsection{Ranking Ensemble Learning Definition}
Let $\mathcal{F}=\{f^1,f^2,...,f^K\}$ be $K$ basic models that are trained for $K$ different objectives~(such as click, buy, and favorite, etc.),
$\mathcal{I}(u,c)=\{i_1,i_2,...,i_N\}$ be the union set of $K$ recommended basic item lists for user $u$ in session environment context $c$~(e.g., time and location), and $S^k_{n}(u,c)=f^k(i_n,u,c)$ be the predicted score given by basic model $k$ on item $n$.
The goal of ranking ensemble learning is to learn a weighted ensemble score $S^{ens}_{n}(u,c)$ for each item $i_n$ in $\mathcal{I}$,
\begin{equation}
    S^{ens}_{n}(u,c) = \sum_{k=1}^{K}w^k_n(u,c)\cdot S^k_n(u,c)
    \label{eq:ens}
\end{equation}

Where $w^k_n(u,c)\in \mathbb{R}$ denotes the weight of the $k$-th basic model for item $i_n$.
The weights are learnable with the help of side information, e.g., user intents and item categories.
The items in $\mathcal{I}$ are sorted according to $S_n^{ens}$, and are compared with a ground truth order of ranking $\pi^{u,c}=\{\pi_1,\pi_2,...,\pi_N\}$, which is sorted to users' interactions with a pre-defined priority of user feedback, e.g., Buy>Click>Examine.
 The priority can be defined by business realities and will not influence the model learning strategy.
The definition of ranking ensemble learning is similar to previous work~\cite{liu2022generalized,balchanowski2022aggregation,oliveira2020rank}, except that we conduct ensemble on heterogeneous basic models with different objectives, which makes the problem more difficult.
The main notations are shown in Table~\ref{tab:notation}.

\begin{table}[]
    \caption{Notations. $u$ and $c$ denote user and context, respectively.}
    \centering
    \small
    \begin{tabular}{ll}
    \hline
     \textbf{Notation}    &  \textbf{Description} \\\hline
      $\mathcal{F}=\{f^1,...,f^K\}$   & Set of $K$ basic models. \\
      $\mathcal{I}(u,c)$   & The union set of items generated with $\mathcal{F}$. \\
      $S^k_n(u,c)$   & Predicted score from basic model $k$ on item $n$. \\
      $z^k_{mn}(u,c)$   & The difference between scores $S^k_n(u,c)-S^k_m(u,c)$  \\
      $w^k_n(u,c)$   & Ensemble weight of item $n$ in basic model $k$. \\
      $S_n^{ens}(u,c)$   & The final ensemble score of item $n$. \\
      $\pi_n(u,c)$ & Ground truth ranking of item $n$. \\
      $\pi^{u,c}$   & List of ground truth, $\pi^{u,c}=\{\pi_1(u,c),...,\pi_n(u,c)\}$. \\
      $Int$ & Distribution of user intent. \\
      $l_m, l_b, l_{p-l}$   & Point-wise, pair-wise and list-wise loss function. \\ 
      A & The ambiguity term in ensemble learning loss.\\
         \hline
    \end{tabular}
    \label{tab:notation}
\end{table}

\subsection{User Intent Definition}
\label{subsec:intent_definition}
When aggregating basic models optimized with different objectives, users' intent about behaviors and item categories are both essential.
Therefore, we define a user's intent in a visit as a probability distribution of item categories and behaviors,
\begin{equation}
    Int \sim P_{int}(I,B), \sum_{I\times B}P_{int}(I,B) = 1
\end{equation}
Where $I$ and $B$ indicate the item category intents and behavior intents, respectively.
The types of categories and behaviors vary with recommendation scenarios.
For instance, in online shopping, $I$ can be product class, and $B$ may include clicking and buying.
In music recommender systems, $I$ may be music genre, while $B$ can contain listening and purchasing albums.
In experiments, user intents $Int$ are predicted from users' historical interactions and environment context.

\subsection{Ranking Losses}
\label{subsec:ranking_losses}
Three representative losses are generally leveraged in the recommendation scenario, namely point-wise, pair-wise, and list-wise loss. 
We will theoretically and empirically illustrate the effectiveness of ranking ensemble with three losses in the following sections.

For a given user $u$ under session context $c$ with multi-level ground truth ranking $\pi^{u,c}=\{\pi_1,\pi_2,...\pi_N\}$ on an item set $\mathcal{I}=\{i_1,i_2,...,i_N\}$, the loss of score list $\mathbf{S}(u,c)=\{S_1,S_2,...,S_N\}$ is defined as~($u$ and $c$ are omitted):
\begin{itemize}
    \item \textbf{Point-wise Loss} As $\pi$ is a multi-level feedback based on a group of user feedback, the Mean Squared Error~(MSE) loss is utilized as a representative point-wise loss,
    \begin{equation}
        l_m(\pi,\mathbf{S}) = \frac{1}{N}\sum_{n=1}^Nl_m(\pi_n,S_n) := \frac{1}{N}\sum_{n=1}^N (S_n-\pi_n)^2
        \label{eq:mse}
    \end{equation}
    \item \textbf{Pair-wise Loss} We leverage the Bayesian Personalized Ranking~(BPR) loss~\cite{rendle2012bpr}. Following the negative sampling strategy for multi-level recommendation~\cite{luo2020spatial}, a random item from one level lower is paired with a positive item at each level,  
    \begin{equation}
    \begin{aligned}
        l_{b}(\pi,\mathbf{S}) :&= 
        \frac{1}{N^+}\sum_{l=1}^L \sum_{n,m\in {I^{+}_l,I^-_l}}l_b(S_n,S_m)\\
        l_b(S_n,S_m) &= -log\sigma(S_n-S_m)
    \end{aligned}
        \label{eq:bpr}
    \end{equation}
    Where $L$ is the number of interaction levels~(e.g., buy, click, and exposure), $N^+$ is the number of positive items of all levels, $I^{+}_l$ and $I^{-}_l$ are positive and one-level-lower negative item set for level $l$, and $\sigma$ is the sigmoid function.
    \item \textbf{List-wise Loss} Following~\cite{liu2022generalized}, we adopt the Plackett-Luce (P-L) model as the likelihood function of ranking predictions,
    \begin{equation}
        P_{p-l}(\pi|\mathbf{S}) = \frac{1}{N}\prod_{n=1}^N\frac{\exp(S_{\pi_n})}{\sum_{m=n}^{N}\exp(S_{\pi_m})}
    \end{equation}
    Where $\pi_n$ indicates the $n$-th item sorted by ground truth $\pi$.
    The corresponding list-wise loss function is
    \begin{equation}
        l_{p-l}(\pi,\mathbf{S}):=-\log[P_{p-l}(\pi|\mathbf{S})] 
        \label{eq:list}
    \end{equation}
\end{itemize}
\section{Theoretical effectiveness of ranking ensemble learning}
\label{sec:theorem}

To prove the effectiveness of our proposed item-level ranking ensemble learning in Eq.\ref{eq:ens}, we aim to prove that the loss of ensemble learning scores 
 $\mathbf{S}^{ens}=\{S_n^{ens}\}$ can be smaller than any of the loss of basic-model scores $\mathbf{S}^k=\{S_n^k\}$ for point-wise, pair-wise, and list-wise loss, i.e. $l(\pi,\mathbf{S}^{ens})\leq\sum_{k=1}^K \mathbf{w}^k\  l(\pi,\mathbf{S}^k),\forall \mathbf{w}^k, l\in \{l_m,l_b,l_{p-l}\}$. 
 In this way, we can claim that there exist some combinations of weights $\mathbf{w}^k$ to achieve results better than all basic models.

Inspired by previous studies in ensemble learning, error ambiguity~(EA) decomposition~\cite{krogh1994neural} provides an upper bound for ensemble loss $l(\pi,\mathbf{S}^{ens})$, which helps conduct the above proof.
For basic lists with loss $\{l(\pi,\mathbf{S}^k)\}$, EA decomposition tries to split ensemble loss $l(\pi,\mathbf{S}^{ens})$ into a weighted sum of basic-model loss~($\sum_kw_k l(\pi,\mathbf{S}^k), \forall w_k$) minus a positive ambiguity term $A$\footnote{The ambiguity $A$ is sometimes called the \textit{diversity} in EA decomposition. We use \textit{ambiguity} to denote it to avoid confusion with the term \textit{item diversity} in recommendation.} of basic models, so that the upper bound of $l(\pi,\mathbf{S}^{ens})$ is controlled by both basic-model losses and ambiguity.
It was recently proved in ranking tasks with the same weights for a basic list~(i.e., $w_n^k=w_m^k,\forall n=m$)~\cite{liu2022generalized}.
However, different weights should be assigned for different items in our setting.
Therefore, we need to verify whether EA decomposition is still available.
To summarize, we try to prove that loss functions can be rewritten as $l(\pi, \mathbf{S}^{ens})\leq \sum_k\sum_n w_n^kl_n^k(\pi, \mathbf{S}^k_n)-A, \forall w_n^k$ for point-wise, pair-wise, list-wise loss in the following.

\subsection{Point-wise Loss}
\begin{theorem}[Generalized EA Decomposition Theory for Point-wise Loss]
Given a set of score lists $\{S_n^k|k\in\{1,2,...,K\},n\in\{1,...,N\}\}$ from $K$ basic models on $N$ items, and a weighted ensemble model $S_n^{ens}=\sum\nolimits_{n=1}^Nw_n^kS_n^k$ with $w_n^k\geq 0$and $\sum\nolimits_{k=1}^Kw_n^k=1$, the MSE loss of the $n$-th ensemble score $S_n^{ens}$ can be decomposed into two parts,
\begin{equation}
l_m(\pi_n,S_n^{ens}) = \sum_{k=1}^{K}w_n^k l_m(\pi_n,S_n^k) - \sum_{k=1}^Kw_n^k A_n^k
\label{eq:mse_decompose}
\end{equation}
where $A_n^k$ indicates the ambiguity term,
\begin{equation}
    A_n^k=(S_n^k-S_n^{ens})^2
    \label{eq:diversity_mse}
\end{equation}
\end{theorem}
\vspace{-0.3cm}
\begin{proof}
For each basic-model score $S_n^k$, we expand the MSE loss $l_m(\pi_n,S_n^k)$ in Eq.~\ref{eq:mse} around point $S_n^{ens}$ by Taylor expansion with Lagrange type reminder ($\pi_n$ is removed when the meaning is clear),
\begin{equation}
    l_m(S_n^k) = l_m(S_n^{ens}) + \frac{\partial{l_m(\tilde{S}_n^{ens})}}{\partial \tilde{S}_n^{ens}}(S_n^{k}-S_n^{ens}) + \frac{1}{2!} \frac{\partial^2l_m(\tilde{S}_n^{ens})}{(\partial \tilde{S}_n^{ens})^2}(S_n^k-S_n^{ens})^2
\end{equation}
Where $\tilde{S}_n^{ens}$ is an interpolation point between $S_n^{ens}$ and $S_n^k$.
Define $A_n^k$ as Eq.\ref{eq:diversity_mse}, we weighted sum losses of all basic models as follows,
\begin{equation}
\begin{split}
    \sum_{k=1}^K w_n^k l_m(S_n^k) =& \sum_{k=1}^K [w_n^k l_m(S_n^{ens}) + \frac{\partial{l_m(\tilde{S}_n^{ens})}}{\partial \tilde{S}_n^{ens}} w_n^k(S_n^{k}-S_n^{ens})+ w_n^kA_n^k] \\
    =&l_m(S_n^{ens})+\frac{\partial{l_m(\tilde{S}_n^{ens})}}{\partial \tilde{S}_n^{ens}}\left(\sum_{k=1}^K w_n^k S_n^{k}-S_n^{ens}\right) + \sum_{k=1}^Kw_n^kA_n^k \\
    =& l_m(S_n^{ens})+ \sum_{k=1}^Kw_n^kA_n^k
\end{split}
\end{equation}
The first equation is due to $\sum_{k=1}^K w_n^k=1$ and $\partial^2 l_m/(\partial S)^2=2$, and the second equation is due to Eq.~\ref{eq:ens}.
Therefore,
\begin{equation}
    l_m(S_n^{ens}) = \sum_{k=1}^K w_n^k l_m(S_n^k) - \sum_{k=1}^K w_n^k A_n^k
\end{equation}
Proof done.
\end{proof}

Since the ambiguity $A_n^k$ in Eq.~\ref{eq:diversity_mse} is positive and $w_n^k\geq 0$, Eq.~\ref{eq:mse_decompose} follows the form of EA decomposition.
Ranking ensemble with item-level weights for point-wise loss is effective theoretically, since the ensemble loss is smaller than weighted sum of basic model losses with any $w_n^k$ as long as $w_n^k\geq 0$ and $\sum_{k=1}^K w_n^k=1$.

For brevity, we will omit statements of score lists and ensemble formulas in the following theorems.

\subsection{Pair-wise Loss}
\begin{theorem}[Generalized EA Decomposition Theory for Pair-wise Loss]
    When $w_n^k\geq 0$, $\sum_{k=1}^Kw_n^k=1$, and $|w_m^k-w_n^k|\leq \delta, \forall m,n$,
    the BPR loss of a pair of ensemble scores $S_n^{ens}$ and $S_m^{ens}$ can be decomposed into
\begin{equation}
    l_b(S_n^{ens},S_m^{ens}) <\sum_{k=1}^K w_n^k l_b(S_n^k,S_m^k) + \delta \sum_{k=1}^{K}S_m^k - \sum_{k=1}^Kw_n^k A_{nm}^k
    \label{eq:bpr_decompose}
\end{equation}
Where $A_{nm}^k$ is the ambiguity of scores generated from basic models,
\begin{equation}
    A_{nm}^k = \sigma(\tilde{z}^{ens})(1-\sigma(\tilde{z}^{ens}))\sum_{k=1}^K w_n^k(z_{nm}^k-z_{nm}^{ens})^2
    \label{eq:a_bpr}
\end{equation}
$z_{nm}^{*}=S_{n}^*-S_m^*$ denotes the differences between scores.
\end{theorem}

Due to space limitation, we only show key steps in the proof:

\begin{proof}
Let $z_{nm}^k=S_n^k-S_m^k$ and $l_b(z_{nm}^*)=l_b(S_n^*,S_m^*)$ in Eq.\ref{eq:bpr}, we expand $l_b(z_{nm}^k)$ around $z_{nm}^{ens}$ by Taylor expansion,
\begin{equation}
    \begin{aligned}
    l_b(z_{nm}^k) &= l_b(z_{nm}^{ens})+ \frac{\partial{l_b(\tilde{z}_{nm}^{ens})}}{\partial \tilde{z}_{nm}^{ens}}(z_{nm}^k-z_{nm}^{ens}) + \frac{1}{2!}\frac{\partial^2l_b(\tilde{z}_{nm}^{ens})}{(\partial \tilde{z}_{nm}^{ens})^2}\\
    &:= l_b(z_{nm}^{ens})-B_{nm}^k+A_{nm}^{k}
    \end{aligned}
    \label{eq:proof2_1}
\end{equation}
Where $\tilde{z}_{nm}^{ens}$ is an interpolation point between $z_{nm}^{ens}$ and $z_{nm}^k$.
With the limitation that $\sum_{k=1}^K w_n^k=1$ and $|w_n^k-w_m^k|\leq \delta$, the weighted sum of $B_{nm}^k$ is limited by 
\begin{equation}
\begin{split} 
    \sum_{k=1}^K w_n^k B_{nm}^k &= [1-\sigma (z_{nm}^{ens})]\sum_{k=1}^K w_n^k (z_{nm}^k -z_{nm}^{ens}) \\
    &\le \sum_{k=1}^K |w_n^k-w_m^k| |S_m^k| \leq \sigma \sum_{k=1}^K S_m^k
    \label{eq:app_bpr_b}
\end{split}
\end{equation}
Sum both sides of Eq.\ref{eq:proof2_1} with weights, we get
\begin{align}
    l_b(z_{nm}^{ens}) < \sum_{k=1}^K w_n^k l_b(z_{nm}^k) + \sigma \sum_{k=1}^K S_m^k - \sum_{k=1}^K w_n^k A_{nm}^k
    \label{eq:app_bpr}
\end{align}

Proof done.
\end{proof}

The range limitation of $w_n^k$ leads to $\delta\leq 1$. And in pair-wise loss, the order rather than the values of scores matters. So the second term in Eq.~\ref{eq:bpr_decompose}~($\delta \sum_{k=1}^K S_m^k < \sum_{k=1}^K S_m^k$) can be arbitrarily small. Meanwhile, the ambiguity $A_{nm}^k$ is semi-positive. Therefore, Eq.\ref{eq:bpr_decompose} follows the form of EA decomposition, and our ranking ensemble method with pair-wise loss is effective theoretically.

\subsection{List-wise Loss}

\begin{theorem}[Generalized EA Decomposition Theory for List-wise Loss] 
When $w_n^k\geq 0$, $\sum_{k=1}^Kw_n^k=1$, and $|w_n^k-w_m^k|\leq \delta$ for any $m$ and $n$, the list-wise loss of ensemble scores $\mathbf{S}^{ens}=\{S_1^{ens},S_2^{ens},...,S_n^{ens}\}$~(sorted with $\pi$) can be decomposed as
\begin{equation}
    l_{p-l}(\pi,\mathbf{S}^{ens}) < \sum_{k=1}^K w^k_{\max} l_{p-l}(\pi,\mathbf{S}^{k}) + \delta N S_{\text{sum}}^{\max} - \sum_{k=1}^K\sum_{n=1}^N w^k_n A_n^k
    \label{eq:list_decompose}
\end{equation}
Where $w^k_{\max}$ denotes the maximum of all weights in list $k$, $A_n^k$ is the ambiguity at position $n$,
\begin{equation}
    A_n^k = \frac{\left[\sum_{m=n+1}^{N}\exp(-\tilde{z}_{nm}^{ens})(z_{nm}^k-z_{nm}^{ens})\right]^2}{\left(1+\sum_{m=n+1}^{N}\exp(\tilde{z}_{nm}^{ens})\right)^2}
    \label{eq:a_list}
\end{equation}
$S_{\text{sum}}^{\max}$ is defined as
\begin{equation}
    S_{\text{sum}}^{\max} = \max_{m = 1}^N \sum_{k = 1}^K S_m^k
    \label{eq:S_sum_max}
\end{equation}
    $z_{nm}^*=S_n^*-S_m^*$ denotes the differences between scores.
\end{theorem}

Due to space limitation, we only show key steps in the proof:

\begin{proof}
    
We define the score difference $ z_{n:N}=[z_{n+1},...,z_N]=[S_n-S_{n+1},...,S_n-S_{N}]$ and the logarithm pseudo-sigmoid function,
\begin{equation}
    g_n(z_{n:N}) = \log\left(1+\sum_{m=n+1}^N \exp(-z_{nm})\right)
\end{equation}
For each basic model of a list of items $\mathbf{S}^k=\{S_n^k|n\in\{1,2,...,N\}\}$, the PL loss is $l_{p-l}(\mathbf{S}^k)=\sum_{n=1}^N g_n(z^k_{n:N})$.
We expand $g_n(z^k_{n:N})$ around point $z^{ens}_{n:N}$ by Taylor expansion with Lagrange type reminder,
\begin{equation}
    \begin{aligned}
        g_n(z^k_{n:N}) =& g_n(z^{ens}_{n:N}) + [\nabla g_n( z^{ens}_{n:N})]^T [z^k_{n:N}-z^{ens}_{n:N}] \\
        &+ \frac{1}{2!}[z^k_{n:N}-z^{ens}_{n:N}]^T H_n(\tilde{z}^{ens}_{n:N})[z^k_{n:N}-z^{ens}_{n:N}] \\
        :=& g_n(z^{ens}_{n:N}) - B_n^k + A_n^k
    \end{aligned}
\end{equation}
With the limitation that $\sum_{k=1}^Kw_n^k=1$, $w_n^k\geq 0$, and $|w_n^k-w_m^k|<\delta, \forall n,m$,
the weighted sum of $B_n^k$ on $K$ basic models will be
\begin{equation}
    \begin{aligned}
        \sum_{k=1}^K w_n^k B_n^k &= \frac{\sum_{m=n+1}^N \exp(-\tilde{z}_{nm}^{ens}) \sum_{k=1}^K w_n^k[z_{nm}^k-z_{nm}^{ens}]}{1+\sum_{m=n+1}^N \exp(-\tilde{z}_{nm}^{ens})} \\
        &< \delta \cdot S_{\text{sum}}^{\max}
    \end{aligned}
    \label{eq:app_pl_b}
\end{equation}
Therefore, sum from $n=1$ to $n=N$, we get
\begin{equation}
    l_{p-l}(\mathbf{S}^{ens})<\sum_{k=1}^K w_{max}^k l_{p-l}(\mathbf{S}^k) + \delta N S_{\text{sum}}^{\max} - \sum_{n=1}^N\sum_{k=1}^K w_n^k A_n^k
    \label{eq:app_pl}
\end{equation}
Proof done.
\end{proof}

Because in the list-wise optimization, the order rather than the values of scores matters, $S_n^{ens}$ can be arbitrarily small. Meanwhile, the ambiguity term $A_n^k$ is semi-positive.
Therefore, Eq.\ref{eq:list_decompose} conforms to the EA decomposition theory, and our ranking ensemble method with list-wise loss is effective.

\subsection{Ensemble Loss for Model Training}

The above theorems guarantee our proposed ranking ensemble learning method in theory for three representative loss functions.
With EA decomposition theory, we prove that the loss of ensemble list is smaller than any weighted sum combination of losses of basic lists: $l_{ens}(\pi,S^{ens})\leq \sum_k w_k l_k(\pi,S^{k})-A + \Delta, \forall w_k\leq 0,\sum_{k=1}^K w_k=1$, where $A$ is a positive ambiguity term, and $\Delta$ is arbitrarily small.
Therefore, the ensemble loss $l_{ens}(\pi,S^{ens})$~(i.e., differences between ensemble list and ground truth) is possible to be smaller than any basic list loss with suitable weights $\{w_n^k\}$, and larger ambiguity $A$ will lead to a smaller bound of ensemble loss. Thus, it can be effective for our ranking ensemble task.

In practice, since basic lists are fixed~(so $l_k(\pi,S^{k})$ are constants), we aim to minimize the ensemble ranking loss $l_{ens}(\pi,S^{ens})$ and maximize the ambiguity $A$. Therefore, the loss function for ranking ensemble learning, $l_{el}$, is defined as follows,
\begin{equation}
    l_{el} = l_{ens}(\pi,\mathbf{S}^{ens}) - \alpha A
    \label{eq:final_loss}
\end{equation}
where $l_{ens}(\pi,\mathbf{S}^{ens})$ can be any of the $l_m(\pi,\mathbf{S}^{ens})$, $l_b(\pi,\mathbf{S}^{ens})$, and $l_{p-l}(\pi,\mathbf{S}^{ens})$, and $A$ indicates the ambiguity term.
For BPR and P-L loss, there exists an interpolation $\tilde{S}_n^{ens}=S_n^{ens}+\theta (S_n^k-S_n^{ens})$ in $A$.
To simplify the calculation, we let $\theta \rightarrow 0$ without loss of generality.

\section{Intent-aware Ranking Ensemble Method}

\subsection{Overall Framework}

\begin{figure}
    \centering
    \includegraphics[width=0.9\linewidth]{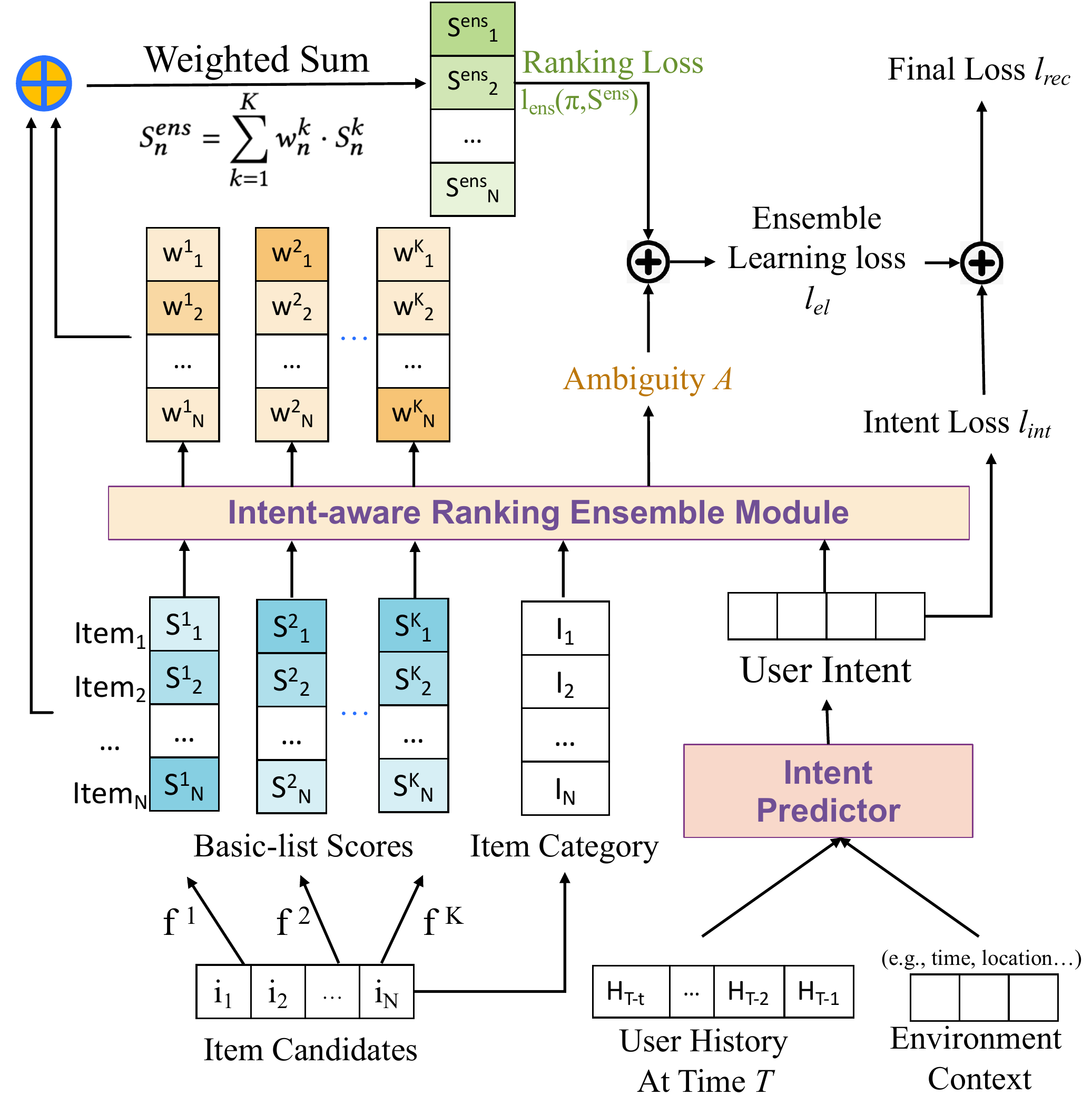}
    \caption{Overall framework of the Intent-aware Ranking Ensemble Learning~(IntEL) model, where $l_{ens}$ and $l_{rec}$ are generated by Eq.\ref{eq:final_loss} and Eq.\ref{eq:joint_loss}, respectively.}
\label{fig:framework}
\end{figure}

After we proved the effectiveness of item-level weights $\{w_n^k\}$ for ranking ensemble with three different loss functions, we need to design a neural network for learning the weights $w_n^k$.
As shown in Section~\ref{subsec:intent_definition}, users' intents about behaviors and item categories help aggregate the basic lists, while intents are not available in advance.
Therefore, an intent predictor and an intent-aware ranking ensemble module are designed for our method.

The main framework of our Intent-aware Ensemble Learning~(IntEL) method is shown in Figure~\ref{fig:framework}.
For a user $u$ at time $T$, 
user intents $Int$ are predicted with an intent predictor from her historical interactions and current environment context.
Then, with $N$ candidate items generated from $K$ basic models, an intent-aware ranking ensemble module is adopted to integrate basic list scores, item categories, and the predicted user intents. 
The output of the ensemble module is item-level weights $\{w_n^k\}$ for each item $n$ and basic model $k$.
Eventually, weighted sum of all basic list scores constructs the ensemble scores $\{S^{ens}_k\}$ for a final list.
Since we focus on the ranking ensemble learning problem, a straight-forward sequential model is used for intent prediction in Section~\ref{subsec:intent_predictor}, and we pay more attention to the design of ensemble module in Section~\ref{subsec:model_ensemble}.
IntEL is optimized with a combination of ranking loss $l_{ens}(\pi,S)$, ambiguity loss $A$, and intent prediction loss $l_{int}$.
Details about the model learning strategy will be discussed in Section~\ref{subsec:training_strategy}.

\subsection{User Intent Predictor}
\label{subsec:intent_predictor}

As defined in Section~\ref{subsec:intent_definition}, user intent describes a multi-dimensional probability distribution $Int$ over different item categories and behaviors at each user visit.
The goal of the intent predictor is to generate an intent probability distribution for each user visit.
We predict intents with users' historical interactions and environment context, as both historical habits and current situations will influence users' intents.

For a user $u$ at time $T$, her historical interactions from $T-t$ to $T-1$ and environment context~(such as timestamp and location) at $T$ are adopted to predict her intent at $T$, where $t$ is a pre-defined time window.
Enviroment context is encoded into embedding $c(u,T)$ with a linear context encoder.
Two sequential encoders are utilized to model historical interactions at user visit~(i.e., session) level and item level.
Session-level history helps learn users' habits about past intents, while item-level interactions express preferences about item categories in detail. 
At the session level, the intents and context of each historical session are embedded with two linear encoders, respectively.
Then two embeddings are concatenated and encoded with a sequential encoder to form an embedding $h_s(u,T)$.
At the item level, ``intent'' of each positive historical interaction can also be represented by its behavior type and item category. Then item-level ``intent''s are embedded with the same intent encoder as session-level, and fed into a sequential encoder to form item-level history $h_i(u,T)$. 
The sequential encoder can be any sequential model, such as GRU~\cite{chung2014empirical}, transformer~\cite{sun2019bert4rec}, etc.
Finally, context $c(u,T)$, session-level $h_s(u,T)$, and item-level $h_i(u,T)$ are concated for a linear layer to predict intent $\hat{Int}(u,T)$~($u$ and $T$ are omitted),
\begin{equation}
    \hat{Int} = \mathrm{Softmax}(\mathbf{W}^{I}[c,h_s,h_i]+b^{I})
    \label{eq:intent_prediction}
\end{equation}
Where $\mathbf{W}^{I}$ and $b^{I}$ are linear mapping parameters.

\begin{figure}
    \centering
    \includegraphics[width=0.95\linewidth]{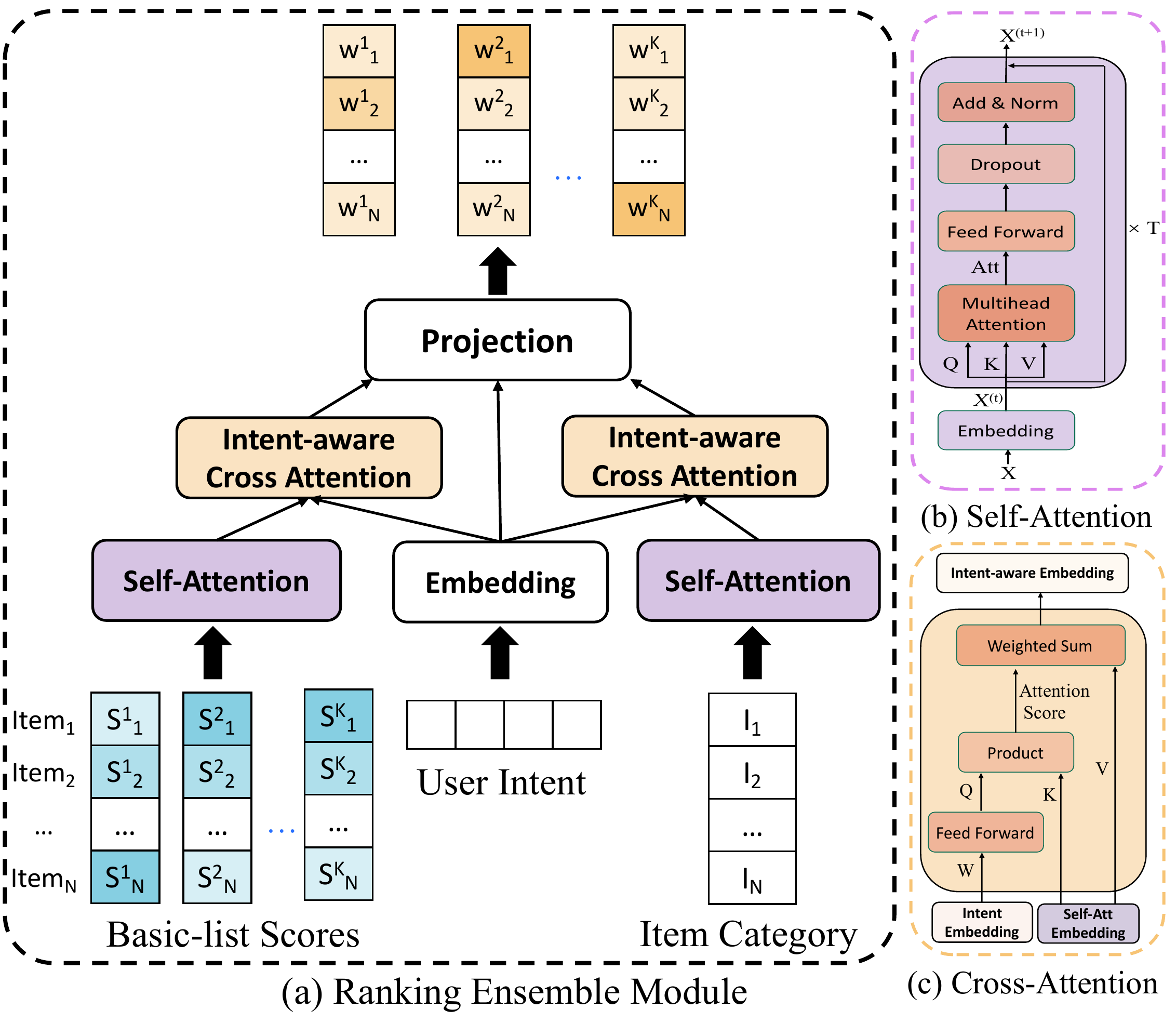}
    \caption{Structure of the intent-aware ranking ensemble module.}
    \label{fig:ensemble_module}
\end{figure}

\subsection{Design of Ensemble Module}
\label{subsec:model_ensemble}

The structure of the intent-aware ranking ensemble module is shown in Figure~\ref{fig:ensemble_module}(a).
Since the final weights $\{w_n^k\}$ should be learned from both behavior and item categories aware of user intents, the predicted intents, single-behavior-objective basic list scores, and categories of items in basic lists are adopted as inputs.

Firstly, lists of item scores $\{S_n^k|k\in\{1,2,...,K\},n\in\{1,2,...,N\}\}$ are fed into a self-attention layer to represent the relationship among item scores in the same basic list. 
Item categories $\{I_n|n\in\{1,2,...,N\}\}$ are also encoded with a self-attention layer to capture the intra-list category distributions. 
The self-attention structure consists of a linear layer to embed scores $\{S_n^k\}$~(or categories $\{I_n\}$) into $d_e$-dimensional representations $\mathbf{S}\in \mathbb{R}^{N \times d_e}$ (or $\mathbf{I}\in \mathbb{R}^{N \times d_e}$) and $T$ layers of multi-head attentions, which follow the cross-relation attention layer proposed by \citet{wang2020toward}, as shown in Figure~\ref{fig:ensemble_module}(b).

Secondly, user intent $Int$ is embedded into $d_{int}$ dimension with a linear projection $Int_d=\mathbf{W}^{i} Int \in \mathbb{R}^{N \times d_{int}}$.
Then the influences of user intent on representations of scores and features are obtained with cross-attention layers,
\begin{align}
    \mathbf{A}_s &= \mathrm{Attention}(Q=\mathbf{W}^QInt_d,K=\mathbf{S},V=\mathbf{S}) \\
    \mathbf{A}_i &= \mathrm{Attention}(Q=\mathbf{W}^QInt_d,K=\mathbf{I},V=\mathbf{I}) \label{eq:item_attention}
\end{align}
Where the projection matrix $\mathbf{W}_Q \in \mathbb{R}^{d_e\times d_{int}}$ is shared between two intent-aware attention modules.
Since behavior intents and category intents are associated when users interact with recommenders, we use the holistic user intents to guide the aggregations of both basic list scores and item categories rather than splitting the intents into two parts.  

Finally, weights $\{w_n^k\}$ should be generated from all information.
Intent-aware score embeddings $\mathbf{A}_s$, intent-aware item category embeddings $\mathbf{A}_i$, and intent embedding $Int_d$ are concatenated and projected into space of $\mathbb{R}^K$ to get the weight matrix $\mathbf{W}\in \mathbb{R}^{N\times K}$,

\begin{equation}
    \mathbf{W} = \mathbf{W}^w \cdot ([\mathbf{A}_s,\mathbf{A}_i,Int_d])
\end{equation}
Where $\mathbf{W}^w\in \mathbb{R}^{K\times (2d_e+d_{int})}$ is a trainable projection matrix.
The output matrix $\mathbf{W}=\{w_n^k\}$ is used as the weights for summing basic model scores as in Eq.~\ref{eq:ens}. 

\subsection{Model Learning Strategy}
\label{subsec:training_strategy}

Since an end-to-end framework is to train the intent predictor module and intent-aware ranking ensemble module, joint learning of two modules is utilized for model optimization. 

To optimize ranking ensemble results
according to theorems via EA decomposition in Section~\ref{sec:theorem}, 
ensemble learning loss $l_{el}$ consists of $l_{ens}(\pi,\mathbf{S}^{ens})$ and $A$ as in Eq.\ref{eq:final_loss}.
Meanwhile, accurate user intents will guide the ranking ensemble, so an intent prediction loss is also used for model training.
Since user intents are described by multi-dimensional distributions, KL-divergence~\cite{csiszar1975divergence} loss $l_{int}$ is adopted to measure the distance between true intents $Int$ and predicted intents $\hat{Int}$.
The final recommendation loss $l_{rec}$ is a weighted sum
\begin{equation}
    l_{rec} = l_{el} + \gamma l_{int} = l_{ens}(\pi,\mathbf{S}^{ens}) - \alpha A + \gamma l_{int}
    \label{eq:joint_loss}
\end{equation}
Where $l_{ens}(\pi,\mathbf{S}^{ens})$ is the ranking ensemble loss, $A$ is the ambiguity term, and $l_{int}$ is the intent prediction loss.
$\alpha$ and $\gamma$ are hyper-parameters to adjust the weights of ambiguity and intent loss, respectively.

\section{Experiments}

\subsection{Experimental Setup}
\subsubsection{Dataset}
Experiments are conducted on a public online shopping recommendation dataset \textbf{Tmall}\footnote{https://tianchi.aliyun.com/dataset/dataDetail?dataId=42} and a private local life service recommendation dataset \textbf{LifeData}. 

\begin{table}[]
    \centering
    \caption{Datasets statistics in ranking ensemble experiments.}
    \label{tab:dataset}
    \small
    \begin{tabular}{lllllll}
    \toprule
Dataset & \#Item & \#User & \#Click & \#Buy & \#Fav. & \#Session \\
\midrule
Tmall & 142.5k & 148.4k & 2.439m & 185.5k & 271.9k & 743.3k \\
LifeData & 165.5k & 82.7k & 819.4k & 82.5k & - & 559.2k\\
\bottomrule
\end{tabular}
\end{table}

\textbf{Tmall} is a multi-behavior dataset from the IJCAI-15 competition, which contains  half-year user logs with Click, Add-to-favorite~(Fav.), and Buy interactions on Tmall online shopping platform.
We employ the data in September for ensemble learning and exclude items and users with less than 3 positive interactions.
Following the data processing strategy by~\citet{shen2021multi}, we treat 
a user's interactions within a day as a session.
Three-week interactions before the ensemble dataset are used for the generation of basic-model scores, which will be discussed in Section~\ref{subsubsec:base_model}.

\textbf{LifeData} comes from a local life service App, where the recommender provides users with nearby businesses such as restaurants or hotels. Users may click or buy items on the platform.
One-month interactions of a group of anonymous users are sampled, and users and items with less than 3 positive interactions are filtered.
A user's each visit~(i.e., entering the App) is defined as a session, and sessions with positive interactions are retained.

Basic models are optimized for each of the behaviors, which will be introduced in Section~\ref{subsubsec:base_model}.
Ranking ensemble is conducted at session level, and interactions in most $t=20$ historical sessions are considered in the intent predictor.
Detailed statistics are shown in Table~\ref{tab:dataset}, which includes the dataset for ensemble learning only while excluding the data used for basic model generation.
Moreover, training data for basic models have no overlap with ensemble learning data.

\subsubsection{Basic-model Score 
 and Intent Generation}
\label{subsubsec:base_model}
In IntEL, basic scores are pre-generated and fixed during ranking ensemble.
For \textbf{Tmall}, we adopted DeepFM~\cite{guo2017deepfm} as basic models to train three models with Click, Fav., and Buy objectives separately.
In each session, we select the top 30 items predicted by each basic model to construct three item sets, and take the union of them, plus positive interactions of the session, to form the basic item lists for reranking.
Please refer to our public repository for details about basic model training strategy\footnotemark[3].
For \textbf{LifeData}, two basic score lists are used for ranking ensemble,
which are sorted by predicted clicking probability and buying probability provided by the platform, respectively.

As for intents, 
in \textbf{Tmall}, $|B|=3$, and we merge categories with less than 50 items, resulting in category $|I|=357$. 
In \textbf{LifeData}, $|I|=6$ and $|B|=2$.
Hence, the dimension for intent $Int$ is 1071 for \textbf{Tmall} and 12 for \textbf{LifeData}.
Intent ground truth $Int$ probability is calculated from all positive interactions in each session.

\subsubsection{Baseline Methods}

We compare IntEL against basic models and several ranking ensemble baselines as follows,

1. \textbf{Single $XXX$}: Use one of the basic models' scores to rank the item list. $XXX$ indicates Click, Fav., and Buy, respectively.

2. \textbf{RRA}~\cite{kolde2012robust}: An unsupervised ensemble method, where items are sorted with their significance in basic-model lists.

3. \textbf{Borda}~\cite{borda1784memoire}: An unsupervised ensemble method to take the average ranks of all basic models as the final ranking.

4. \textbf{$\lambda$Rank}~\cite{burges2006learning}: A gradient-based optimization method used for learning2rank task. We regard items as documents, basic-model scores and item categories as document features, and MLP as a backbone model.

5. \textbf{ERA}~\cite{oliveira2016evolutionary}: An evolutionary method to aggregate some basic-list features with Genetic Algorithm~(GA), where fitness function is calculated on validation set.

6. \textbf{aWELv}~\cite{liu2022generalized}: A personalized ranking ensemble method to assign weights at basic model level, i.e., $w_n^k=w_m^k$ for any $n,m$. We adopt the list-wise training loss following~\cite{liu2022generalized}.

7. \textbf{aWELv+Int/IntEL}: Two variations of \textbf{aWELv} considering user intents. Intents are predicted as a feature for aWELv+Int. The IntEL module is used for predicting list-level weights for aWELv+IntEL.

Our methods are shown as \textbf{IntEL-MSE}, \textbf{IntEL-BPR}, and \textbf{IntEL-PL} with three different kinds of loss functions.

\subsubsection{Experimental settings}
\label{subsubsec:exp_evaluation}
We split both datasets along time: the last week is the test set, and the last three days from the training set is the validation set.
The priority for the mutli-level ground truth $\pi$ are Buy>Favorite>Click>Examine for \textbf{Tmall}, and Buy>Click>Examine for \textbf{LifeData}.
As for evaluation, we adopt NDCG@3, 5, and 10 to evaluate the ensemble list $S^{ens}$ on the multi-level ground truth $\pi$~(i.e., all) and each behavior objective.

We implement IntEL model in \textit{PyTorch}, and the code of IntEL and all baselines are released\footnote[3]{https://github.com/JiayuLi-997/IntEL-SIGIR2023.}.
Each experiment is repeated with 5 different random seeds and average results are reported.
All models are trained with Adam until convergence with a maximum of 100 epochs.
For a fair comparison, the batch size is set to 512 for all models.
We tuned the parameters of all methods over the validation set, where the learning rate are tuned in the range of $[1e-4, 1e-2]$ and all embedding size are tuned in $\{16,32,64\}$.
Specifically, for IntEL, we found that it has stable performance when GRU~\cite{chung2014empirical} with embedding=128 is used for the intent predictor, and self-attention with $T=2$ for Tmall and $T=1$ for LifeData.
The ambiguity loss weight $\alpha$ is set to 1e-5, 1e-5, and 1e-4 for IntEL-MSE, IntEL-BPR, and IntEL-PL. 
Hyper-parameter details are released\footnotemark[3].

\subsection{Overall Performance}
\begin{table}[]
\small
    \centering
    \caption{Main differences between two datasets. Pos. indicates positive interactions.}
    \label{tab:dataset_diff}
    \begin{tabular}{r|ccc}
    \toprule
Dataset & \#Intent & Avg. Session Length & Avg. Pos./Session \\
\midrule
Tmall & 1,071 & 68.37 & 3.73 \\
LifeData & 12 & 32.78 & 1.47 \\
\bottomrule
\end{tabular}
\end{table}

\begin{table*}
\caption{Results of IntEL with three different loss functions and baseline methods on Tmall. Boldface shows the best result. Underline indicates the best baseline. Notation **/* demonstrates significantly better than the best baseline with p<0.05/0.01.}
\label{tab:overall_performance_tmall}
\small
\begin{tabular}{r|ccc|ccc|ccc|ccc}
\toprule
 & \multicolumn{3}{c|}{All-NDCG@K} & \multicolumn{3}{c|}{Click-NDCG@K} & \multicolumn{3}{c|}{Fav.-NDCG@K} & \multicolumn{3}{c}{Buy-NDCG@K}\\
\multirow{-2}{*}{Model} & K=3 & K=5 & K=10 & K=3 & K=5 & K=10& K=3 & K=5 &K=10& K=3 & K=5&K=10 \\
\midrule
Single Click & 0.1356 & 0.1473 & 0.1673 & 0.1435 & 0.1532 & 0.1721 &0.0701 & 0.0829 & 0.1014 & 0.0918&0.1057 &0.1243 \\
Single Fav. & 0.0752 & 0.0874 & 0.1066 & 0.0779 & 0.0894 & 0.1083 & 0.0630 & 0.0748 & 0.0920 & 0.0492 & 0.0607 & 0.0765 \\
Single Buy & 0.0572 & 0.0689 & 0.087 & 0.0587 & 0.0699 & 0.0878 & 0.0393 & 0.0489 & 0.0638 & 0.0632 & 0.0776 & 0.0974 \\ 
\midrule
RRA & 0.0960 & 0.1093 & 0.1317 & 0.0998 & 0.1120 & 0.1341 & 0.0683 & 0.0813 & 0.1014 & 0.0753 & 0.0907 &0.1122  \\
Borda & 0.1258 & 0.1398 & 0.1626 & 0.1317 & 0.1440 & 0.1660 & 0.0741 & 0.0880 & 0.1081 & 0.0830 & 0.0989 & 0.1218 \\
$\lambda$Rank & 0.2742 & 0.2797 & 0.3003 & \underline{0.3104} & \underline{0.3064} & \underline{0.3189} & 0.1878 & 0.2122 & 0.2472 & 0.1376 & 0.1586 & 0.1913  \\
ERA & \underline{0.3325} & \underline{0.3378} & \underline{0.3623} & 0.2301 & 0.2420 & 0.2716 & \underline{0.1933} & \underline{0.2156} & \underline{0.2502} & \underline{0.1921} & \underline{0.2163} & \underline{0.2504} \\ 
\midrule
aWELv & 0.1387 & 0.1533 & 0.1770 & 0.1469 & 0.1584 & 0.1811 & 0.0837 & 0.0986 & 0.1197 & 0.1025 & 0.1198 & 0.1436 \\
aWELv+Int & 0.1398 & 0.1574 & 0.1784 & 0.1484 & 0.1592 & 0.1822 & 0.0903 & 0.1016 & 0.1183 & 0.1030 & 0.1259 & 0.1445 \\
aWELv+IntEL & 0.1427 & 0.1556 & 0.1774 & 0.1535 & 0.1620 & 0.1821 & 0.0906 & 0.0934 & 0.1120 & 0.1042 & 0.1263 & 0.1451  \\
\midrule
IntEL-MSE & \textbf{0.4257}** & \textbf{0.4364}** & \textbf{0.4676}** & \textbf{0.4693}** & \textbf{0.4680}** & \textbf{0.4712}** & \textbf{0.2943}** & \textbf{0.3271}** & \textbf{0.3731}** & \textbf{0.2433}* & \textbf{0.2760}** & \textbf{0.3100}** \\ 
IntEL-BPR & 0.3992* & 0.3859* & 0.3755 & 0.4417** & 0.4157** & 0.3960* & 0.2791** & 0.2943** & 0.3068* & 0.2344* & 0.2508* & 0.2630 \\
IntEL-PL & 0.4041** & 0.3865* & 0.3678 & 0.4367** & 0.4060** & 0.3829** & 0.2811** & 0.2934** & 0.3032* & 0.2355* & 0.2472* & 0.2594 \\

\bottomrule
\end{tabular}
    
\end{table*}

\begin{table*}
\caption{Results of IntEL with three different loss functions and baseline methods on LifeData. Boldface shows the best result. Underline indicates the best baseline. Notation **/* demonstrates significantly better than the best baseline with p<0.05/0.01.}
\label{tab:overall_performance_private}
\small
\begin{tabular}{r|ccc|ccc|ccc}
\toprule
 & \multicolumn{3}{c|}{All-NDCG@K} & \multicolumn{3}{c|}{Click-NDCG@K} & \multicolumn{3}{c}{Buy-NDCG@K} \\
\multirow{-2}{*}{Model} & K=3 & K=5 & K=10 & K=3 & K=5 & K=10 & K=3 & K=5 & K=10 \\
\midrule
Single Click & 0.4004 & 0.4443 & 0.4972 & 0.4009 & 0.4449 & 0.4980 & 0.6365 & 0.6665 & 0.6918 \\
Single Buy & 0.3102 & 0.3526 & 0.4070 & 0.3102 & 0.3528 & 0.4072 & 0.6893 & 0.7211 & 0.7438 \\
\midrule
RRA & 0.3539 & 0.4020 & 0.4586 & 0.3540 & 0.4022 & 0.4590 & 0.6556 & 0.6865 & 0.7104 \\
Borda & 0.4094 & 0.4538 & 0.5061 & 0.4097 & 0.4541 & 0.5066 & 0.7030 & 0.7250 & 0.7447 \\
$\lambda$Rank & 0.4129 & 0.4487 & 0.4830 & 0.4133 & 0.4492 & 0.4835 & 0.6866 & 0.7083 & 0.7225 \\
ERA & 0.4063 & 0.4451 & 0.5053 & 0.4181 & 0.4579 & 0.5112 & 0.5782 & 0.6307 & 0.6764 \\
\midrule
aWELv & 0.4074 & 0.4531 & 0.5033 & 0.4077 & 0.4535 & 0.5041 & \underline{0.7063} & \underline{0.7339} &  0.7466 \\
aWELv+Int & 0.4150 & 0.4607 & 0.5143 & 0.4151 & 0.4610 & 0.5147 & 0.6962 & 0.7271 & 0.7482 \\
aWELv+IntEL & \underline{0.4174} &\underline{0.4663} & \underline{0.5176} & \underline{0.4189} & \underline{0.4638} & \underline{0.5171} & 0.7036 & 0.7318 & \underline{0.7503} \\
\midrule
IntEL-MSE & 0.4253** & 0.4695* & 0.5211** & 0.4257** & 0.4700** & 0.5217* & 0.7096 & 0.7379 & 0.7498 \\
IntEL-BPR & 0.4308** & 0.4752** & 0.5268** & 0.4312** & 0.4757** & 0.5275** & \textbf{0.7115*} & \textbf{0.7390*} & \textbf{0.7609**} \\
IntEL-PL & \textbf{0.4378**} & \textbf{0.4819**} & \textbf{0.5332**} & \textbf{0.4382**} & \textbf{0.4825**} & \textbf{0.5339**} & 0.7093 & 0.7382* & 0.7604**\\
\bottomrule
\end{tabular}
    
\end{table*}

The overall performances on \textbf{Tmall} and \textbf{LifeData} are shown in Table~\ref{tab:overall_performance_tmall} and Table~\ref{tab:overall_performance_private}, respectively.
We divide all models into four parts: 
The first part evaluates on each single-objective basic model's scores. 
The second is unpersonalized baseline ensemble models, and the third contains personalized baselines: aWELv and its two variants with user intents.
The last part shows our method IntEL with three loss functions.
From the results, we have several observations:

First, our proposed IntEL achieves the best performance on all behavior objectives in both datasets.
IntEL with three loss functions, i.e., IntEL-MSE, IntEL-BPR, and IntEL-PL, outperform the best baselines on most metrics significantly.
Although the two datasets are quite different, as shown in Table~\ref{tab:dataset_diff}, IntEL has stable, great ensemble results on both datasets.

Second, IntEL with different loss functions show different performances on two datasets.
On {Tmall}, IntEL-MSE is better than IntEL-BPR and IntEL-PL. It is because there is four-level ground truth $\pi$~(three behaviors), and
ranking on such diverse item lists is close to rating prediction.
Therefore, IntEL-MSE, which directly optimizes the ensemble scores, performs better than IntEL-BPR and IntEL-PL, which optimize the comparison between rankings.
On {LifeData}, IntEL-PL and IntEL-BPR perform better since {LifeData} has shorter sessions with fewer positive interactions~(as in Table~\ref{tab:dataset_diff}).
So comparison-based BPR and P-L achieve better performance.

Third, comparing different baselines, we find that supervised methods~($\lambda$Rank, ERA, and aWELv) outperform unsupervised RRA and Borda greatly on {Tmall}.
It is because heterogeneous single-behavior objective models~(Single XXX) have diverse performance, 
making rank aggregation difficult for unsupervised methods.

Lastly, aWELv and its variants perform well on {LifeData} but not on {Tmall} since session lists are generally longer~(Table~\ref{tab:dataset_diff}) for Tmall, and list-level weights of aWELv miss useful intra-list information.
So item-level weights that consider item category intents are necessary.
Nevertheless, aWELv is better than basic models in both datasets, which is consistent with the theory.
Moreover, aWELv+Int/IntEL outperform aWELv on most metrics, indicating that user intents contributes to ranking ensemble learning.

\subsection{Further Analysis}

To further explore the performance of our ranking ensemble learning method, we conduct an ablation study, analysis of user intents, and hyper-parameters analysis on the best model for each dataset, i.e., IntEL-MSE for \textbf{Tmall} and IntEL-PL for \textbf{LifeData}.

\subsubsection{Ablation Study}

\begin{figure}
    \centering
    \includegraphics[width=0.95\linewidth]{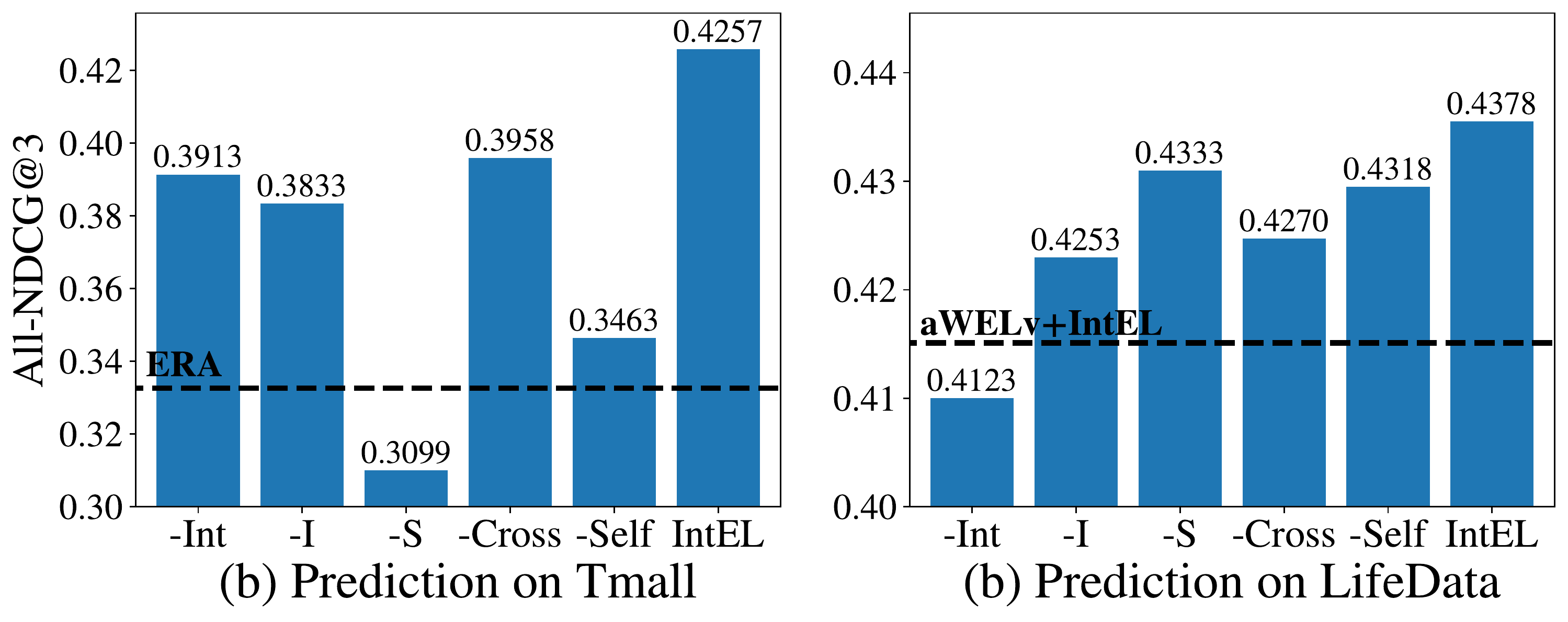}
    \caption{Ablation study. Performance comparison between IntEL and its variant, i.e., without: Intent modeling~(-Int), item categories~(-I), basic score lists~(-S), cross-attention~(-Cross), and self-attention~(-Self).}
    \label{fig:ablation}
\end{figure}
The main contributions of our proposed IntEL include adopting user intents for heterogeneous ranking ensemble and integration of basic-list scores, item categories, and user intents.
We compare IntEL with five variants:
Excluding one of the inputs: \textbf{-Int}~(without intent), \textbf{-I}~(without item categories), and \textbf{-S}~(without basic-list scores).
Replacing two main elements: \textbf{-Cross}, removing the intent-aware cross-attention layer; and \textbf{-Self}, replacing the self-attention layer with a direct connection.

NDCG@3 on the general multi-level ranking list of variants and IntEL are shown in Figure~\ref{fig:ablation}.
Ranking performance drops on all five variants, indicating all inputs and two attention layers contribute to the performance improvement of IntEL.
Removing scores and the self-attention layer both lead to considerable performance decreases on Tmall, showing that intra-list basic scores information is eseential, as long sessions are included in Tmall.
Removing user intents leads to the most dramatic degradation on LifeData, which suggests it is important to adopt user intents for the multiple-objective ranking ensemble.
Nevertheless, the ablation variants still outperform all basic lists~(i.e., Single XXX), which aligns with our proof of loss reduction via EA ambiguity decomposition.

\subsubsection{Influence of User Intent}
\label{subsubsec:intent_accuracy}
\begin{table}[]
\small
    \caption{Intent prediction and ranking ensemble performance comparison with different treatments on user intents.}
    \label{tab:intent}
    \centering
    \begin{tabular}{r|ccc|ccc}
    \toprule
Dataset & \multicolumn{3}{c|}{Tmall} & \multicolumn{3}{c}{LifeData} \\
\midrule
Model & -Int & His.Avg. & IntEL & -Int & His.Avg. & IntEL \\
\midrule
I-Perform & - & 0.1829 & 0.2347 & - & 0.2663 & 0.3298 \\
E-NDCG@3 & 0.3913 & 0.4011 & 0.4257 & 0.4123 & 0.4265 & 0.4378\\
\bottomrule
\end{tabular}
\end{table}

Since intents are essential for ranking ensemble learning,
we explore the influence of intent prediction accuracy.
Therefore, we compare IntEL with two variants: \textbf{-Int}, IntEL without user intents as input, and \textbf{His.Avg.}, predicting a user's current intents as her average historical session intents.

Since \textbf{Tmall} has 1071 intents and \textbf{LifeData} has 12 intents, we utilize NDCG@10 and Macro-F1 as intent performance~(I-perform) indicators, respectively. And ensemble results~(E-NDCG@3) are evaluated by All-NDCG@3 for both datasets.
The results are shown in Table~\ref{tab:intent}.
It indicates that better performance of ranking ensemble is achieved by adding intent prediction and improving prediction accuracy.
Therefore, predicting user intents is helpful for ranking ensemble in recommendation.
On the other, $His.Avg.$ works better than all baselines, providing an efficient and effective possible implementation in application.

\subsubsection{Hyper-parameters Analysis}

\begin{figure}
    \centering
\subcaptionbox{Tmall dataset}{\includegraphics[width=0.4\textwidth]{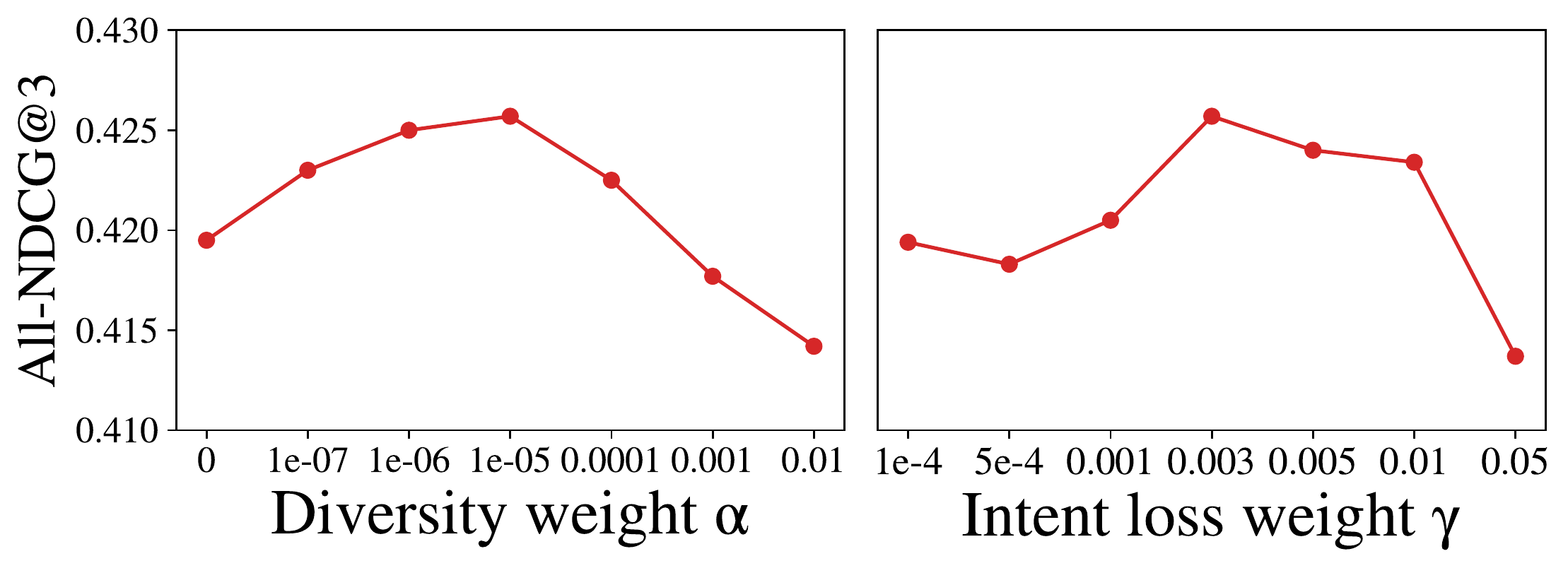}}
\subcaptionbox{LifeData dataset}{\includegraphics[width=0.4\textwidth]{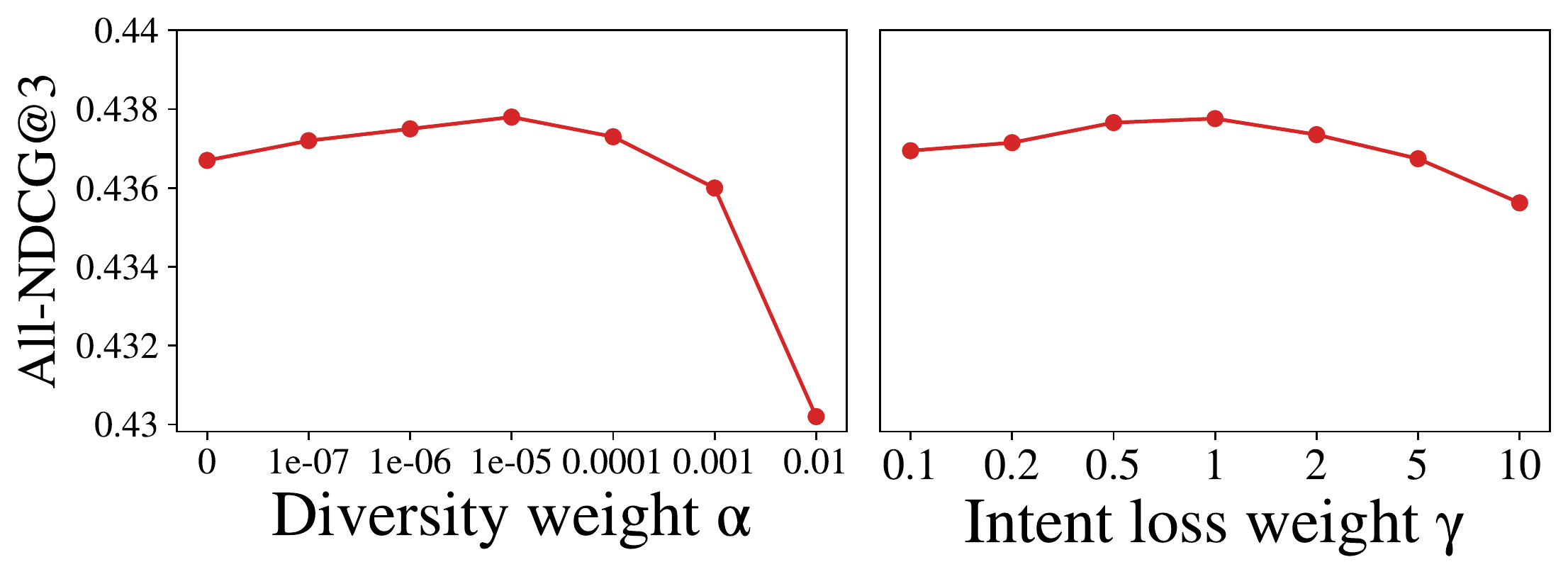}}
    
    \caption{Ranking ensemble results of IntEL with different hyper-parameters.}
    \label{fig:hyper}
\end{figure}

Since the construction of loss $l_{rec}$~(Eq.\ref{eq:joint_loss}) is essential for our method, we analyze the influence of hyper-parameters during model optimization.
Two hyper-parameters are considered in the optimization loss $l_{rec}$: $\alpha$, the weight for basic list ambiguity $A$; $\gamma$, the weight for intent prediction loss $l_{int}$.
All-NDCG@3 with different hyper-parameters on two datasets are shown in Figure~\ref{fig:hyper}.
It illustrates that too large or small $\alpha$ will both lead to ranking ensemble performance decrease. Especially when $\alpha$ is too large, the model will focus on maximizing basic model ambiguity $A$ to minimize $l_{rec}$, while ensemble learning loss $l_{el}$ is less optimized.
As for the intent loss weight $\gamma$, performance on Tmall shows fluctuation with $\gamma$, while performance on LifeData is relatively stable.
It is because intent prediction difficulty differs on two datasets: Tmall contains 1071 types of intents, which is hard to predict accurately, so a proper intent loss weight is essential for predictor optimization, while LifeData has only 12 intents, which is easier to capture and model.

\section{Conclusion}

In this paper, we propose a novel ranking ensemble method IntEL for intent-aware single-objective ranking lists aggregation.
To our knowledge, we are the first to generalize ranking ensemble learning with item-level weights on heterogeneous item lists.
And we are also the first to integrate user intents into rank aggregation in recommendation. 
We generalize the ranking ensemble with item-level weights and prove its effectiveness with three representative loss functions via error-ambiguity decomposition theory.
Based on the proof, we design an ensemble learning loss $l_{el}$ to minimize ranking ensemble loss $l_{ens}$ and maximize ambiguity $A$.
Then we design an intent-aware ranking ensemble learning model, IntEL, to learn weights for heterogeneous lists' ensemble.
In IntEL, a sequential intent predictor and a two-layer attention intent-aware ensemble module are adopted for learning the personalized and adaptive ensemble weights with user intents.
Experiments on two large-scale datasets show that IntEL gains significant improvements on multiple optimization objectives simultaneously.

This study still has some limitations. For basic list generation, we only applied one classical method, DeepFM, for different behaviors separately. However, multi-behavior methods are also possible models to generate multiple basic lists simultaneously, which may lead to different performance for IntEL.
Also, a straight-forward method was adopted to predict intents and incorporate intent prediction loss.
In the future, we will investigate the possibility of integrating more heterogeneous basic lists for other objectives in recommendation with IntEL.
As we find that more accurate user intents will lead to better ranking ensemble performance, we will also try to design more sophisticated intent predictors to achieve better results.

\begin{acks}
We sincerely thank our anonymous reviewers for their insightful feedback. 
This work is supported by the Natural Science Foundation of China (Grant No. U21B2026, 62002191) and Beijing Academy of Artificial Intelligence.
\end{acks}
\newpage
\bibliographystyle{ACM-Reference-Format}
\bibliography{reference}

\end{document}